# BIOMEDICAL DATA WAREHOUSES

Jérôme Darmont[*] and Emerson Olivier

ERIC, Université Lumière Lyon 2

5 avenue Pierre Mendès-France

69676 Bron Cedex

France

Phone: +33 478 774 403

Fax: +33 478 772 375

E-mail: jerome.darmont@univ-lyon2.fr, emerson2fr@yahoo.fr

# BIOMEDICAL DATA WAREHOUSES

## INTRODUCTION

With the growing use of new technologies, healthcare is nowadays undergoing significant changes. The development of electronic health records shall indeed help in enforcing personalized, lifetime healthcare and pre-symptomatic treatment, as well as various analyses over populations of patients. Information-based medicine has to exploit medical decision-support systems and requires the analysis of various, heterogeneous data, such as patient records, medical images, biological analysis results… (Saad, 2004)

Data warehousing technologies (Inmon, 2002; Kimball & Ross, 2002) are now considered mature and can form the base of such a decision-support system. Though data warehousing primarily allows the analysis of numerical data, its underlying concepts remain valid for what we term *complex data*. To summarize, data may be qualified as complex if they are (Darmont et al., 2005):

- *multiformat*, i.e., represented in various formats (databases, texts, images, sounds, videos...);

and/or

- *multistructure*, i.e., diversely structured (relational databases, XML document repositories, file collection...);

and/or

- *multisource*, i.e., originating from several different sources (distributed databases, the Web...);

and/or

- *multimodal*, i.e., described through several channels or points of view (radiographies and audio diagnosis of a physician, data expressed in different scales or languages...);

and/or

- *multiversion*, i.e., changing in terms of definition or value (temporal databases, periodical surveys with evolving items...).

In this context, the warehouse measures, though not necessarily numerical, remain the indicators for analysis, which is still performed following different perspectives represented by dimensions. Large data volumes and their dating are other arguments in favor of this approach (Darmont et al., 2003). Data warehousing can also support various types of analysis, such as statistical reporting, on-line analysis (OLAP) and data mining.

The aim of this article is to present an overview of the existing biomedical data warehouses and to discuss the issues and future trends in this area. We illustrate this topic by presenting the design of an innovative, complex data warehouse for personal, anticipative medicine.

BACKGROUND

The first family of biomedical data warehouses we identify are repositories tailored for supporting data mining (Prather et al., 1997; Tchounikine et al., 2001; Miquel & Tchounikine, 2002; Sun et al., 2004). However, since data mining techniques take "attribute-value" tables as input, some of these warehouses (Prather et al., 1997; Sun et al., 2004) do not bear the usual multidimensional, star-like architecture. This modeling choice precludes OLAP navigation, and is not very evolutionary: new analysis axes, or dimensions, cannot be easily plugged into the warehouse.

The most "canonical" medical data warehouse among these proposals is a cardiology warehouse (Tchounikine et al., 2001; Miquel & Tchounikine, 2002). Its aim is to ease medical data mining by integrating data and processes into a single warehouse. However, raw sensor data (e.g., electrocardiograms) are stored separately from multidimensional data (e.g., patient identity, therapeutic data), while it might be interesting to integrate them all.

A second family is constituted of data warehouses that focus on molecular biology and genetics (Schönbach et al., 2000; Engström & Asthorsso, 2003; Eriksson & Tsuritani, 2003; Sun et al., 2004; Shah et al., 2005), and bear interesting characteristics. For instance, some of them include metadata and ontologies from various public sources such as RefSeq or Medline (Engström & Asthorsso, 2003; Shah et al., 2005). The incremental maintenance and evolution of the warehouse is also addressed (Engström & Asthorsso, 2003). However, the particular focus of these approaches makes them inappropriate to more general needs, which may be both different and much more diversified.

Eventually, Boussaïd et al. (2006) recently proposed an XML (eXtended Markup Language)-based methodology, named X-Warehousing, for warehousing complex data. This approach has been applied on a corpus of patient records extracted from the Digital Database for Screening Mammography[1]. The warehouse is a collection of XML documents representing OLAP facts that describe suspect areas in mammographies. It aims at breast cancer computer-aided diagnosis.

A COMPLEX DATA WAREHOUSE FOR PERSONALIZED, ANTICIPATIVE MEDECINE

Context and Motivation

Dr Jean-Marcel Ferret, former physician of the French national soccer team, is the promoter of the ==personalized and anticipative medicine== project. His aim is to extend results and empirical advances achieved for high-level athletes to other populations and to make the analyzed subjects the managers of their own health capital, by issuing recommendations regarding, e.g., life style, nutrition, or physical activity. This is meant to support personalized, anticipative medicine. In order to achieve personalized, lifetime healthcare and pre-symptomatic treatment, a decision-support system must allow transverse analyses of given populations of patients and the storage of global medical data such as biometrical, biological, cardio-vascular, clinical, and psychological data. It must also be evolutionary to take into account future advances in medical research. More precisely, such a system must be able to store complex medical data, and allow quite different kinds of analyses to support:

1. personalized and anticipative medicine (in opposition to curative medicine) for well-identified patients;
2. broad-band statistical studies over given populations of patients.

We selected a data warehousing approach to answer this need. A data warehouse can indeed support statistical reporting and cross-analyses along several dimensions. Furthermore, dated, personal data can be stored to propose full patient records to physician users. Data complexity must be handled, though. For instance, multimedia documents such as echocardiograms must be stored and explicitly related to more classical medical data such as the corresponding patient or diagnoses. Users must be able to display and exploit such relationships, either manually (which is currently the case) or automatically (we anticipate the advances of multimedia mining and the development of advanced OLAP operators). The existing research proposals from the literature do not fulfill this requirement. In particular, Tchounikine et al. (2001) store raw sensor data separately from multidimensional data, only as an archive. Finally, Dr Ferret had quite an immediate need for an operational, efficient system. Hence, we chose to rely on the efficacy of a relational implementation. Though XML warehousing is particularly appropriate to complex, medical data, the performance of native XML Database Management Systems (DBMSs) is indeed currently not satisfactory for warehousing purposes, both in terms of storage capacity and response time.

---

[1] http://marathon.csee.usf.edu/Mammography/Database.html

In the remainder of this section, we present the global architecture of our medical data warehouse, two examples of simple datamarts (i.e., of datamarts that store "simple" data), an example of complex datamart (i.e., storing complex data), as well as implementation issues.

Global Architecture

Our data warehouse is organized as a collection of interconnected datamarts sharing common dimensions. Each datamart stores the data related to a given medical field (e.g., biological analysis results, biometrical data, cardio-vascular data, psychological data...). These data are described by dimensions. Some are common to all the datamarts (e.g., patient information, dates), to some of them, or are specific to a given datamart. The warehoused data originate from diverse providers and are very heterogeneous (qualitative or numerical data, texts, medical images...).

To make our solution evolutionary, we adopted a bus architecture (Kimball and Ross, 2002; Firestone, 2002) for our data warehouse. It is composed of a set of conformed dimensions and standardized definitions of facts. In this framework, the warehoused data related to every medical field we need to take into account represent datamarts that are plugged into the data warehouse bus and receive the dimension and fact tables they need. The union of these datamarts may be viewed as the whole data warehouse (Kimball and Ross, 2002).

Using conformed dimensions, i.e., dimensions that have the same meaning and the same modeling in all the fact tables they are linked to, helps in guaranteeing that datamarts are not "stovepipes" that are independent from one another. Hence, when a new aspect of healthcare (e.g., chiropody) needs to be included into the warehouse, it is easier to plug it into the data warehouse bus.

Figure 1 represents the global architecture of our data warehouse. Straight squares symbolize fact tables, round squares symbolize dimensions, dotted lines embed the different datamarts, and the data warehouse bus bears a gray background. It is constituted by dimensions that are common to several datamarts (and thus fact tables). The main dimensions that are common to all our datamarts are patient, data provider, time, and medical analysis (that regroups several kinds of analyses, see the next section). Of course, some datamarts (such as the cardio-vascular datamart) do have specific dimensions that are not shared.

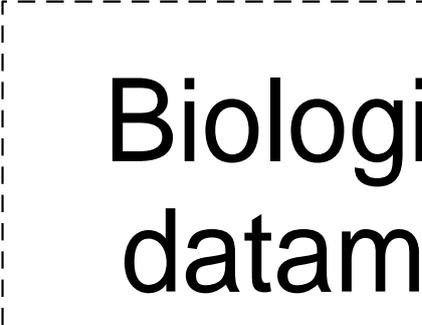

Figure 1: Data warehouse global architecture

Simple Datamarts: the Biological and Biometrical Datamarts

Input biological data are actually the results of various biological examinations (biochemistry, protein checkup, hematology, immunology, toxicology, etc.), which are themselves subdivided into actual analyses (e.g., a hematology examination consists in reticulocyte numbering and a hemogram). These data are available under the form of unnormalized spreadsheet files from different sources. They are thus heterogeneous and often refer to the same examination using different terms or abbreviations, use different units for the same numerical values, etc. This heterogeneity is dealt with during the ETL (Extract, Transform, and Load) process (see the implementation section below).

Biometrical data are measured during medical examinations. They include weight, height, pulse rate, fat percentage, and blood pressure. Though their structure is simpler than that of the biological data, they require a fine granularity that must be taken into account. For exam-ple, the weight of an athlete may be measured twice a day, before and after practice. This has an impact on data warehouse modeling. More precisely, it helps in defining the granularity of the time dimension hierarchy (see below).

Figure 2 represents the architecture of our biological and biometrical datamarts. The biological fact table stores an exam result under the form of a numerical value (e.g., a reticulocyte numbering). It is described by four dimensions: patient, time of the examination, data provider (typically the laboratory performing the analysis), and the analysis itself. The biometrical fact table stores numerical biometrical values (e.g., weight). It is described by the same dimensions than the biological datamart, the "analysis" actually representing a measurement. The patient, data provider, time, and medical examination dimensions are thus all shared, conformed dimensions. Attributes in dimension tables are not detailed due to confidentiality constraints.

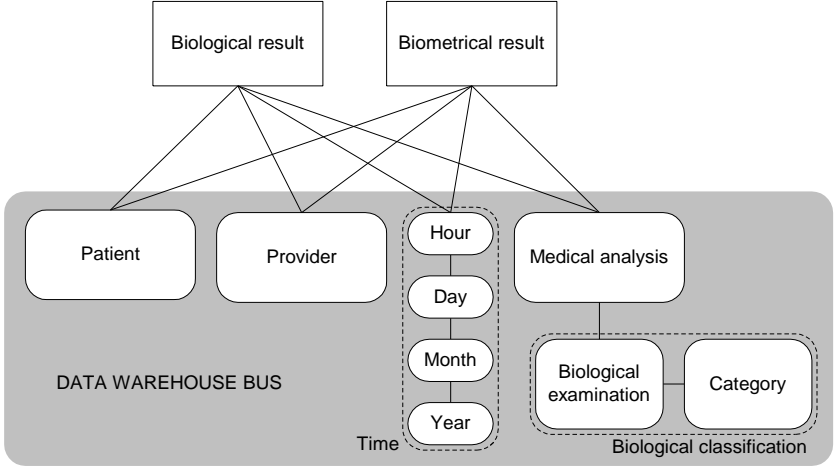

Figure 2: Biological and biometrical datamarts' architecture

Note that the medical analysis and time dimensions are further organized into hierarchies, to take into account the particularities identified into the source data. Here, biological and bio-metrical data are distinguished: the simple biometrical data are not normalized nor organized in a hierarchy, while the biological data are. Hence, the description of biometrical facts only appears in the medical analysis dimension table, while biological facts are further described by a hierarchy of biological examinations and categories.

Each datamart is thus modeled as a snowflake schema, rather than a simpler, classical star schema. Since the biological and biometrical fact tables share their dimensions, our overall data warehouse follows a constellation schema. In such an architecture, it is easy to add in new fact tables described by existing dimensions.

Finally, though this is not indicated on Figure 2, our data warehouse also include metadata that help in managing both the integration of source data into the warehouse (e.g., correspondence between different labels or numerical units, the French SLBC biomedical nomenclature, etc.) and their exploitation during analysis (e.g., the interval between which an examination result is considered normal).

Complex Datamart: the Cardio-vascular Datamart

Figure 3 represents the architecture of our cardio-vascular datamart. The complex nature of source data, which are constituted of raw measurements (e.g., ventricle size), multimedia documents (e.g., echocardiograms) and a conclusion by a physician, cannot be embedded in a single, standard fact table. Hence, we actually exploit a set of interrelated tables that together represent the facts. They are represented as dotted, straight squares on Figure 3.

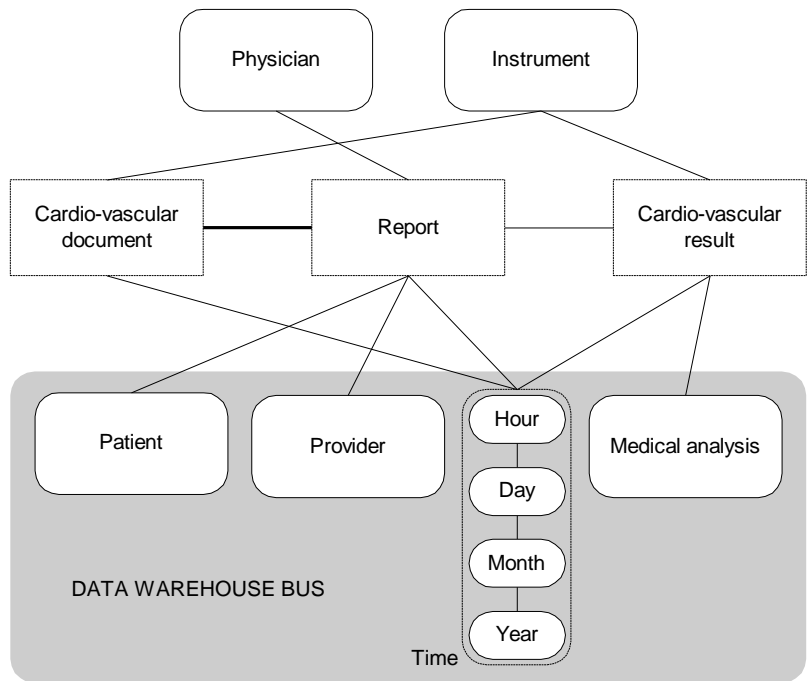

Figure 3: Cardio-vascular datamart's architecture

The report mainly contains the physician's conclusion. It is the central element in our "complex fact". It is linked to several numerical analysis results that help in building the physician's conclusion. It is also related to multimedia documents such as medical images that also help in devising the diagnosis. Note that this many-to-many relationship is represented as a bold line on Figure 3. Some documents may indeed be referred to by several reports, for instance to take into account a patient's evolution over time, through distinct series of echocardiograms. Each component of our "complex fact" may be individually linked to the dimensions. Cardio-vascular documents and results are indeed not descriptors of the would-be fact: report, they are part of a fuzzier fact that is composed of several entities.

Finally, note that cardio-vascular documents cannot currently be exploited by OLAP technologies. However, we need to store them and have them accessible for medico-legal reasons. Furthermore, extensions to the usual OLAP operators should make this exploitation possible in the near future.

Implementation

Our data warehouse is meant to be used on an intranet. Hence, we developed a prototype using popular tools that are freely available, namely the Apache Web server that implements the Secure Sockets Layer (SSL) protocol, which is important for the security of our solution; the MySQL DBMS that is fast and reasonably reliable; and the powerful PHP web scripting language. This software configuration is sufficient for our prototype, very light and easy to deploy, but we can easily switch one component for a more robust one if necessary. This is especially true for the DBMS component, since our data warehouse will have to store voluminous complex data.

At the ETL level, in order to achieve the integration of biological data into our data warehouse (biometrical data are much simpler to integrate), we developed a series of scripts that help in extracting the data from the source spreadsheet files, in transforming them so that they are consistent with the data already present in the warehouse, and eventually in loading them in the data warehouse. This is currently only a semi-automatic process that proceeds in three steps/PHP scripts:

1. remove doubles in the input files;
2. check whether dimension data exist in the data warehouse and create them if not;
3. transform the data whenever necessary (measure units, dimension key correspondence...) and finally load the data into the warehouse.

Eventually, the first analyses we implemented over our data warehouse are divided into two kinds that correspond to the goals we mentioned earlier: patient-oriented results and statistical analyses over groups of patients. For the first kind of output, our front-end tool includes a patient search engine; the display of a given patient's biological and biometrical data following the biological or biometrical examination dimension, respectively; the visual identification of out-of-norm values; and graphical representations of the evolution of any numerical measure over time.

To achieve the second kind of output, we implemented the notion of group of patients. The user can create, modify and suppress such groups (e.g., soccer players, back and forward players, etc.); select group members according to various criteria; and eventually compare the different groups through numerical tables and graphs.

FUTURE TRENDS

The need for medical decision-support systems has nowadays grown very strong, and many such systems have been developed. However, most of them store, process and allow the analysis of "simple" data. Taking complex data into account is now a pressing challenge (Saad, 2004).

We propose in this article a solution for warehousing, i.e., storing complex data. Many research perspectives still lie ahead, though. The multimedia data we store are indeed available for display, but we cannot truly analyze them. Developing OLAP and/or data mining techniques that operate on complex data is currently becoming a research field in itself. We could consider, for instance, medical images as a dimension in a medical warehouse. Aggregating such data with clustering-based OLAP operators (Ben Messaoud et al., 2004) could help in rolling up and drilling down along this dimension.

To achieve this goal, integrating semantic information about the processed complex data is mandatory, and OLAP and data mining techniques must heavily rely on metadata and domain-specific knowledge. The integration of this knowledge into complex data warehouses, and most of all its exploitation for data analysis, are also exciting challenges.

CONCLUSION

We have presented in this paper an overview of biomedical data warehousing, and particularly focused on the complex data warehouse for personal, anticipative medicine we propose. The aim of this tool is to support both patient-oriented analyses and broad-band statistical studies over populations of patients. By adopting a data warehouse bus architecture, we designed our system to be global, to take into account several medical fields, and evolutionary, to take into account future advances in medical research. We have also briefly presented the prototype we implemented to achieve our analysis goals.

The direct perspectives over this work are twofold. The first one concerns the actual contents and significance of the data warehouse. It involves modeling and adding in new datamarts. The psychology datamart from Figure 1 is already implemented and more are currently developed (such as a medical background datamart), being developed, or scheduled. This helps in broadening the scopes of analyses. Other output than statistical reports are also envisaged. Since we adopted a dimensional modeling, OLAP navigation is also definitely possible, and "attribute-value" views could easily be extracted from the data warehouse to allow data mining explorations.

The second kind of perspectives is more technical and aim at improving our prototype. This includes automating and generalizing the ETL process on all the datamarts, which is currently an ongoing task. We also follow other leads to improve the user-friendliness of our interfaces and the security of the whole system, which is particularly primordial when dealing with medical, personal data.

TERMS AND DEFINITION

Information-based medicine: Utilizing information technology to achieve personalized healthcare (Saad, 2004).

Complex data: Data that are not numerical or symbolic (e.g., multimedia, heterogeneous data stored on multiple platforms).

Data warehouse: Subject-oriented, integrated, time-variant and non-volatile collection of data in support of management's decision making process (Inmon, 2002).

Datamart: Logical and physical subset of the overall data warehouse, usually dedicated to a given activity.

Bus architecture: Set of conformed dimensions and standardized definitions of facts (Kimball & Ross, 2002). Datamarts "plug into" this bus to receive the dimensions and facts they need (Firestone, 2002).

OLAP: On-Line Analytical Processing. An approach for processing decision-support, analytical queries that are dimensional in nature.

ETL: Data warehousing process that includes extracting data from external sources, transforming them and finally loading them into the warehouse.